\documentclass[conference, 9pt]{IEEEtran}

\usepackage[square, sort&compress, comma, numbers]{natbib}
\usepackage[colorlinks=true]{hyperref}
\usepackage{amsmath,amssymb,amsfonts}
\usepackage{algorithmic}
\usepackage{graphicx}
\usepackage{textcomp}
\usepackage{pifont}
\usepackage{xcolor}
\usepackage{subcaption}
\usepackage{array}
\usepackage{multirow}
\usepackage{colortbl}

\newlength\bibitemsep
\usepackage[moderate, tracking=normal, title=normal, margins=tight, bibliography=tight, lists=normal, floats=tight, paragraphs=tight, bibbreaks=normal]{savetrees}

\usepackage{titlesec}

\titleformat{\subsubsection}[block] 
{\itshape} 
{\arabic{subsubsection}.} 
{1em} 
{}

\titleformat{\paragraph}[runin] 
{\bfseries} 
{}{5pt}{}[~-- ] 
\titlespacing{\paragraph}{1.5em}{0pt}{0.5em}

\hypersetup{colorlinks,
    linkcolor=ACMRed,
    citecolor=ACMPurple,
    urlcolor=ACMDarkBlue,
    filecolor=ACMDarkBlue}

\RequirePackage{graphicx, xcolor}
\definecolor[named]{ACMBlue}{cmyk}{1,0.1,0,0.1}
\definecolor[named]{ACMYellow}{cmyk}{0,0.16,1,0}
\definecolor[named]{ACMOrange}{cmyk}{0,0.42,1,0.01}
\definecolor[named]{ACMRed}{cmyk}{0,0.90,0.86,0}
\definecolor[named]{ACMLightBlue}{cmyk}{0.49,0.01,0,0}
\definecolor[named]{ACMGreen}{cmyk}{0.20,0,1,0.19}
\definecolor[named]{ACMPurple}{cmyk}{0.55,1,0,0.15}
\definecolor[named]{ACMDarkBlue}{cmyk}{1,0.58,0,0.21}

\def\BibTeX{{\rm B\kern-.05em{\sc i\kern-.025em b}\kern-.08em
    T\kern-.1667em\lower.7ex\hbox{E}\kern-.125emX}}

\begin{document}
\newcolumntype{C}[1]{>{\centering\let\newline\\\arraybackslash\hspace{0pt}}m{#1}}

\title{ForgeBench: A Machine Learning Benchmark Suite and Auto-Generation Framework for Next-Generation HLS Tools \\}

\author{
\IEEEauthorblockN{Andy Wanna$^{1*}$, Hanqiu Chen$^{1*}$, Cong (Callie) Hao$^1$}
\IEEEauthorblockA{$^1$ \textit{School of Electrical and Computer Engineering} - Georgia Institute of Technology,  $^*$\textit{Denotes Equal Contribution}\\
Email: \{awanna3, haniqu.chen, callie.hao\}@gatech.edu}
\vspace{-20pt}
}

\maketitle
\begin{abstract}

Although High-Level Synthesis (HLS) has attracted considerable interest in hardware design, it has not yet become mainstream due to two primary challenges. First, current HLS hardware design benchmarks are outdated as they do not cover modern machine learning (ML) applications, preventing the rigorous development of HLS tools on ML-focused hardware design. Second, existing HLS tools are outdated because they predominantly target individual accelerator designs and lack an architecture-oriented perspective to support common hardware module extraction and reuse, limiting their adaptability and broader applicability. Motivated by these two limitations, we propose ForgeBench, an ML-focused benchmark suite with a hardware design auto-generation framework for next-generation HLS tools. In addition to the auto-generation framework, we provide two ready-to-use benchmark suites. The first contains over 6,000 representative ML HLS designs. We envision future HLS tools being architecture-oriented, capable of automatically identifying common computational modules across designs, and supporting flexible dataflow and control. Accordingly, the second benchmark suite includes ML HLS designs with possible resource sharing manually implemented to highlight the necessity of architecture-oriented design, ensuring it is future-HLS ready. ForgeBench is open-sourced at \url{https://github.com/hchen799/ForgeBench}.

\end{abstract}


\section{Introduction}

\begin{figure*}[!t]
    \centering
    \includegraphics[width=0.95\linewidth]{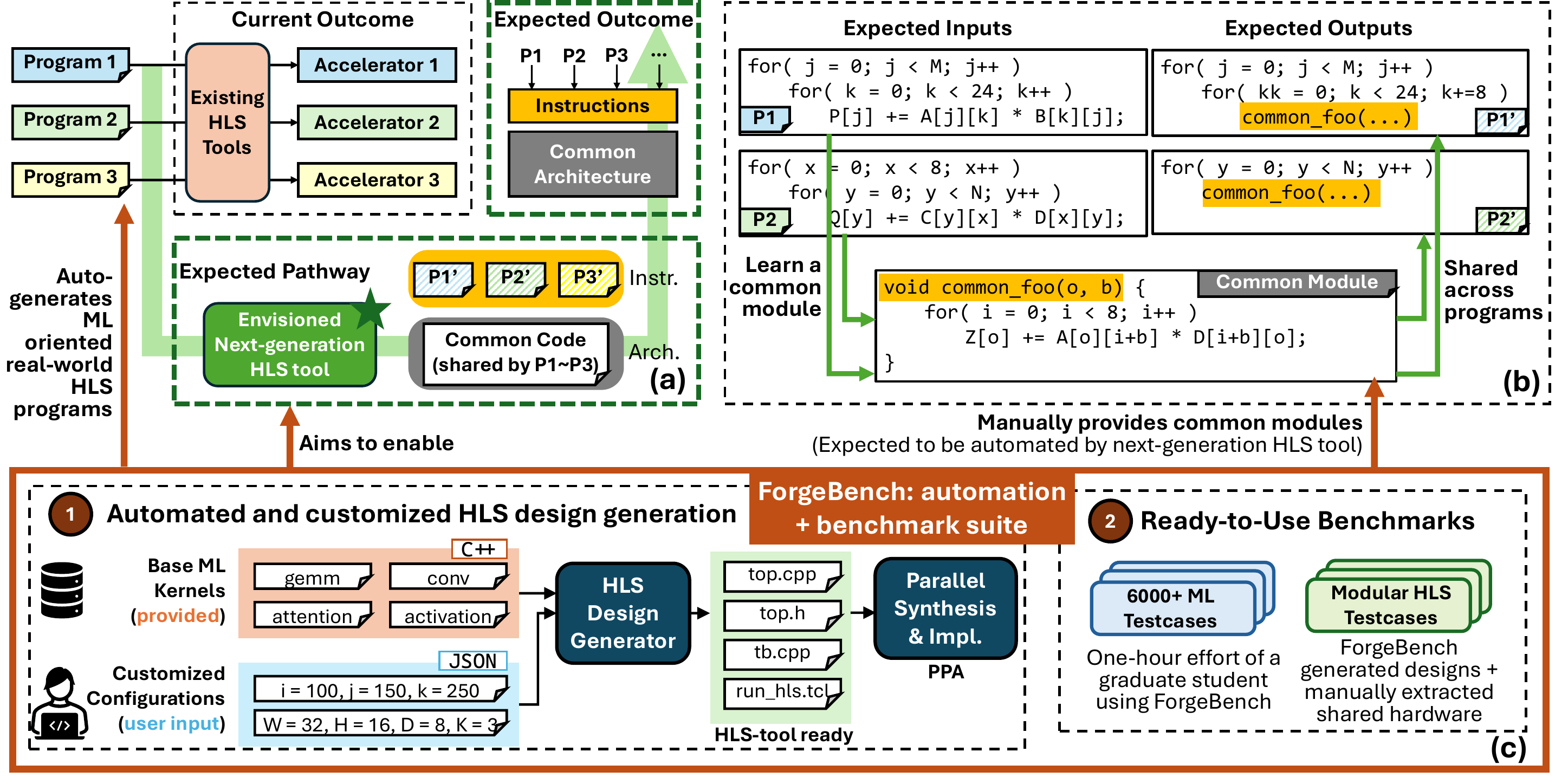}
    \caption{Proposed benchmarking framework enabling future HLS tools for architecture design.}
    \vspace{-15pt}
    \label{fig:workflow}
\end{figure*}

While High-Level Synthesis (HLS) has gained increased traction in hardware design, it has not become the mainstream yet because of two key limitations. \ding{182} \textbf{Outdated HLS benchmarks:} Existing hardware design benchmarks are not ML-ready and HLS-oriented, preventing the effective development of HLS tools for ML hardware. For instance, MachSuite~\cite{MachSuite} and Rosetta~\cite{Rosetta} comprise outdated algorithms and lack coverage of emerging ML models and applications. Other C benchmarks, such as Rodinia~\cite{Rodinia}, designed for GPU benchmarking, and PolyBench~\cite{polybench}, intended primarily for software compilers, have limited applicability to evaluating HLS tools. \ding{183} \textbf{Outdated HLS tools:} Although existing HLS tools~\cite{OverGen, TAPA, HeteroCL, DSAGEN, xilinx_vitis_hls} have significantly improved hardware design efficiency for accelerators, they struggle to design \textit{architectures}. These tools synthesize each input design into a unique accelerator through static dataflow analysis tailored to specific programs. However, with the rapid evolution and diversity of ML applications, new individual hardware accelerators are developed for each new model, leading to significant hardware resource redundancy. This is because different ML models often share substantial computational components. For example, LLaMa~\cite{LLaMA} and GPT~\cite{GPT} can share the computation kernels of attention and General Matrix Multiplication (GEMM), differing only slightly in activation functions. Therefore, we envision that future HLS tools should be capable of automatically identifying and extracting common computational modules across different hardware accelerators for resource sharing, supporting flexible dataflow and control. Part a) of Fig. \ref{fig:workflow} depicts a high-level comparison of the existing HLS tools and how a modular (architecture-oriented) HLS tool may synthesize hardware. An example comparing standard HLS designs to a modularized implementation is highlighted with part b) of Fig. \ref{fig:workflow} with a simple nested loop.

Motivated by these two limitations, in this paper, we propose \textit{ForgeBench}, an ML-focused benchmark suite designed specifically for next-generation HLS tools, together with a user-friendly and highly extensible design auto-generation framework, with details shown in Fig.~\ref{fig:workflow}. Our key contributions are summarized as follows:

\begin{itemize}
    \item An HLS design generation framework enabling users to rapidly generate C/C++ HLS compatible hardware designs and obtain synthesis and implementation reports. The framework is highly extensible, facilitating the integration of new designs into the benchmarking suite.  
    \item A ML-representative benchmark suite, containing over 6,000 different HLS designs, to test and evaluate the current C/HLS tools. 
    \item The design generation framework is future-HLS-ready, including a benchmark suite with manually determined shared modules highlighting the necessity of HLS for architecture design.
\end{itemize}

\section{ForgeBench Overview} \label{sec:method}

ForgeBench is an automated, user-friendly framework that rapidly generates diverse modular HLS designs with minimal manual coding. It incorporates a highly extensible C/C++ template library featuring common ML kernels, enabling quick retrieval and assembly of functional modules. The generated designs serve as robust benchmarks for current and future ML-oriented HLS tools. ForgeBench also provides two ready-to-use benchmarks: \ding{182} over 6,000 ML test cases generated automatically via a provided Python script, \ding{183} modular HLS test cases, including HLS designs automatically generated by the tool and the unified designs with manually extracted shared modules. An overview of ForgeBench is depicted in Fig.~\ref{fig:workflow}.

\subsection{Automatic and Customized Design Generation Workflow}

As shown in part c) of Fig.~\ref{fig:workflow}, ForgeBench provides a Python-based HLS design code generator and supports parallel synthesis and implementation of commercialized HLS flows across multiple threads. It is also integrated with an extensible code library of HLS-oriented C/C++ kernels representing commonly used ML modules, offering customizable code templates for users.

In ForgeBench, users provide customized configurations for each HLS design through JSON files, which can be created manually or generated automatically using our provided Python scripts. Each JSON file corresponds to a distinct HLS design featuring different dataflows and ML module call sequences. In the JSON file, users specify the global BRAM and off-chip DRAM (including names and dimensions), as well as the top-level function interfaces for data communication. Users also define the sequence of ML module calls, detailing inputs, outputs, and specific configurations of each module call. As illustrated in the convolution example in Fig.~\ref{fig:workflow}, module call configurations include loop boundaries, input feature dimensions (height, width, channels), and convolution kernel size. Additionally, the JSON file specifies synthesis configurations, such as clock period, top function name, data types, and the chosen HLS workflow \{CSIM, CO-SIM, Synthesis, Implementation\}.

The HLS design generator takes user-defined configurations from JSON files as inputs, utilizes predefined ML module code templates, and automatically generates complete and HLS-ready C/C++ designs. ForgeBench then performs parallel synthesis and implementation across multiple threads, maximizing computational resource utilization and minimizing the execution time. Upon completion, ForgeBench automatically produces comprehensive performance, power and area (PPA) reports for user analysis. While currently, ForgeBench supports the Vitis HLS flow~\cite{xilinx_vitis_hls} for Xilinx FPGA designs, it can be easily extended to seamlessly support other HLS tool flows.

\subsection{Supported ML Models and Operator Kernels} \label{sec:supported_ops}

We support four different types of operators and three relevant ML models in ForgeBench.

\subsubsection{GEMM \& MLPs}



\paragraph{ML Relevance}
Linear Layers, comprising Generalized Matrix Multiplication (GEMM) operations, are fundamental to ML models ranging from MLPs to DNNs and LLMs. Thus, it is critical to benchmark their performance in HLS tools. The baseline GEMM operation generated by ForgBench is a simple nested `i-j-k' loop. We also include operations for vector-matrix multiplications and a dot product. All these operations can optionally be generated with a bias term. Due to the prominence of GEMM operations, they are rarely implemented as naive nested loops in ML accelerators. Standard optimizations seen in~\cite{DNN_Accelerator,HLS4ML,DNN_codesign,ScaleHLS} include unrolling/tiling to compute the operation in parallel and modifying the loop order for better memory access patterns. Therefore, ForgeBench supports any valid loop re-ordering of `i-j-k', with different unroll parameters for each. Optionally, any multiplication unit can be implemented inline instead of as a function. With these parameters, the benchmark can generate a myriad of vector/matrix multiplications with control over low-level hardware implementations, representative of ML accelerator designs. 

\paragraph{Architecture Oriented HLS}
There is a significant opportunity for hardware reuse across GEMM and vector operations. For example, GEMMs of different dimensions can be computed with a shared tile. Additionally, GEMMs can be composed with simpler vector/matrix multiplications or dot-product units. Reordering the computation loops does not mathematically affect the GEMM operator, enabling them to use the same shared module. This variety of hardware reuse categories forms an ideal test suite to benchmark future modular HLS tools.

\subsubsection{Convolution \& DNNs}

\paragraph{ML Relevance}
Convolution and batch normalization (BatchNorm) are two fundamental modules in DNNs~\cite{ResNet, VGG, MobileNet, EfficientNet}. ForgeBench provides base templates for both, enabling the creation of HLS designs for diverse DNN variants. ForgeBench supports configuring these modules with customizable numbers of channels and dimensions for input/output feature maps. The convolution operation is flexible with variable kernel sizes, padding, and stride, with an optional bias term. ForgeBench also includes a grouped convolution kernel as seen in DNNs such as MobileNet~\cite{MobileNet, MobileNetV2, MobileNetV3}, intended to reduce computational complexity over the traditional.

\paragraph{Architecture Oriented HLS}
DNN hardware designs contain significant opportunities for reusing functional modules. These include tiling strategies for convolution and BatchNorm computations, allowing small kernels to be effectively reused. Additionally, different models can share identical convolution blocks. For example, both ResNet-18 and ResNet-34 incorporate 3×3 convolutions with 64 channels and identical input feature dimensions, enabling the reuse of these functional blocks~\cite{ResNet}.

\subsubsection{Transformers \& LLMs}

\paragraph{ML Relevance}
LLMs are built with transformer architectures, relying on attention, normalization, and embedding encoding operations. These operators can all be configured with different sequence lengths and hidden dimensions of input language tokens in ForgeBench. Modern transformer architectures employ diverse attention mechanisms based on the traditional multi-head attention described in GPT~\cite{GPT}. As a result, multi-head attention in ForgeBench has parameters for the number of heads, number of head groups as seen in LLaMA~\cite{LLaMA}, and an optional variable length sliding window motivated by Mistral~\cite{Mistral}. Additionally, we support both layer normalization (LayerNorm) used in the GPT~\cite{GPT} models and the root mean square normalization (RMSNorm) employed by modern models like LLaMA~\cite{LLaMA}. Furthermore, modern models increasingly use rotary position embeddings (RoPE)~\cite{LLaMA, Mistral, Gemma_1}; hence, a corresponding template is included in the framework.

\paragraph{Architecture Oriented HLS}
There are substantial hardware sharing opportunities across LLM operations. First, the tiling strategy in multi-head attention computation enables the reuse of smaller attention modules with fewer heads to construct larger ones. Second, each attention module can be decomposed into smaller computational modules, such as the GEMM and SoftMax modules, facilitating their reuse across different designs and architectures. Finally, similar to DNNs, various LLMs incorporate common computational blocks. For instance, despite differences in their attention mechanisms, GPT and LLaMA utilize the same feed-forward network, enabling large-scale module reuse.

\subsubsection{General Helper Units}

In addition to the modules discussed above, ForgeBench contains a comprehensive collection of general-purpose modules that can be shared across various ML applications. These include load and store modules for data transfers between BRAM and DRAM, a range of activation function modules, a matrix addition module for residual connections, and element-wise matrix multiplication (Hadamard product) modules for gating networks. Other commonly used modules, such as dropout, max pooling, and average pooling, are also included. Together, these helper modules support building flexible end-to-end ML accelerator designs, enabling the representation of a plethora of models and evaluation of modular architecture-oriented HLS designs.

\subsection{Ready-to-Use Benchmark Suites}


We create two HLS benchmark suites with ForgeBench, demonstrating its convenience and effectiveness at generating relevant ML accelerator test cases, highlighted in Fig. \ref{fig:workflow} c). 

\paragraph{ML Testcases} 
The first is a large-scale general HLS benchmark suite, with over 6000 designs, described in section \ref{sec:ml_results}. This suite is created by tuning the configuration parameters of base ML designs. Python scripts were used to explore $\sim$2000 unique configurations for each test suite and generate the JSON config files for ForgeBench, each taking 1 PhD student around 1 hour to write. The JSON files and the generated HLS designs are publicly available. The scripts are also open-source and serve as a baseline for users to write and generate test cases specific to their tools.

\paragraph{Modular HLS Testcases}
 \begin{figure}[!t]
 \centering
     \begin{subfigure}[b]{0.44\linewidth}
         \centering
         \includegraphics[width=\linewidth]{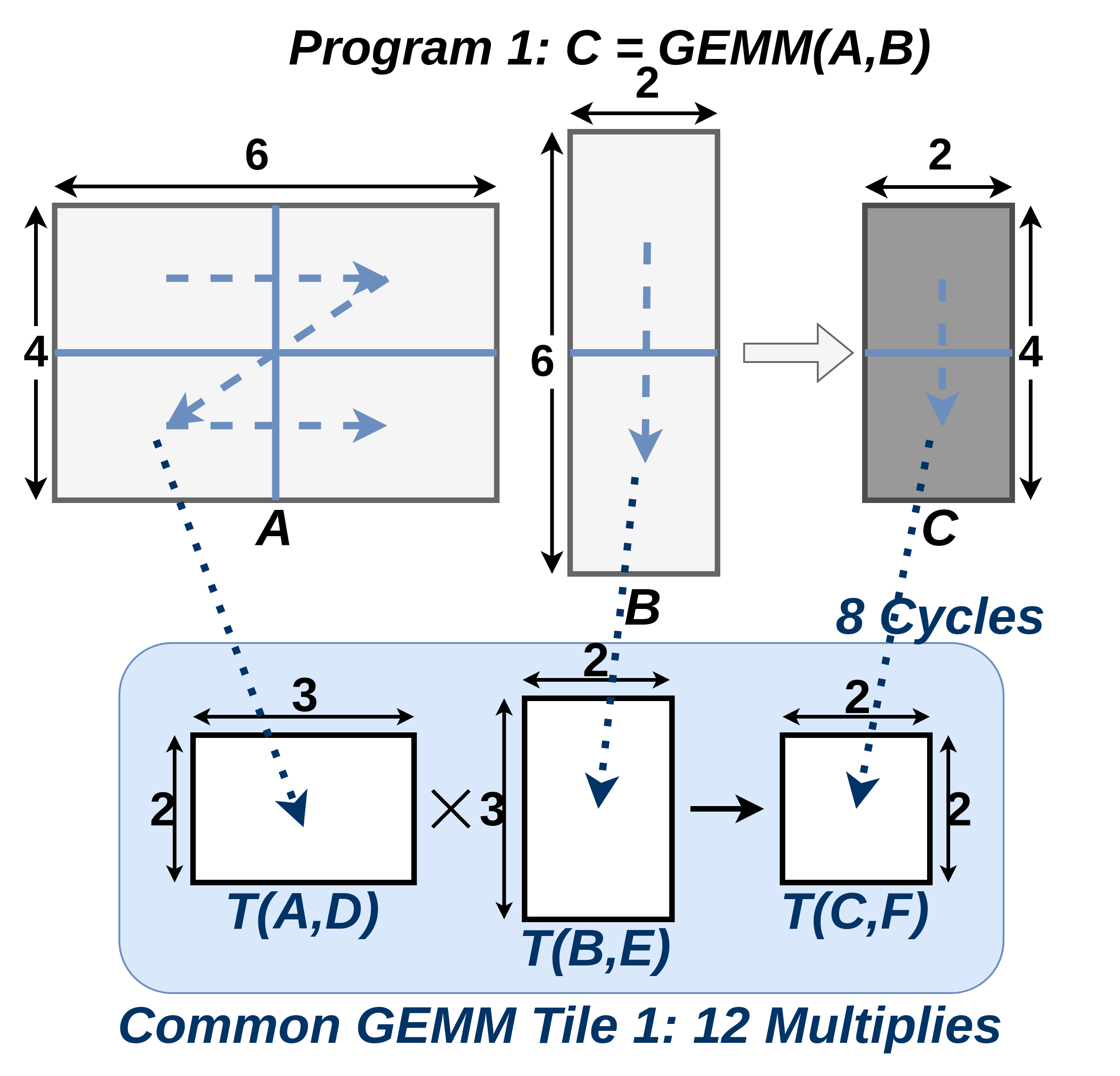}
         \caption{\centering Minimum Tiling GEMM(A,B)}
         \label{fig:tile_p1_t1}
     \end{subfigure}
     \hfill
     \begin{subfigure}[b]{0.44\linewidth}
         \centering
         \includegraphics[width=\linewidth]{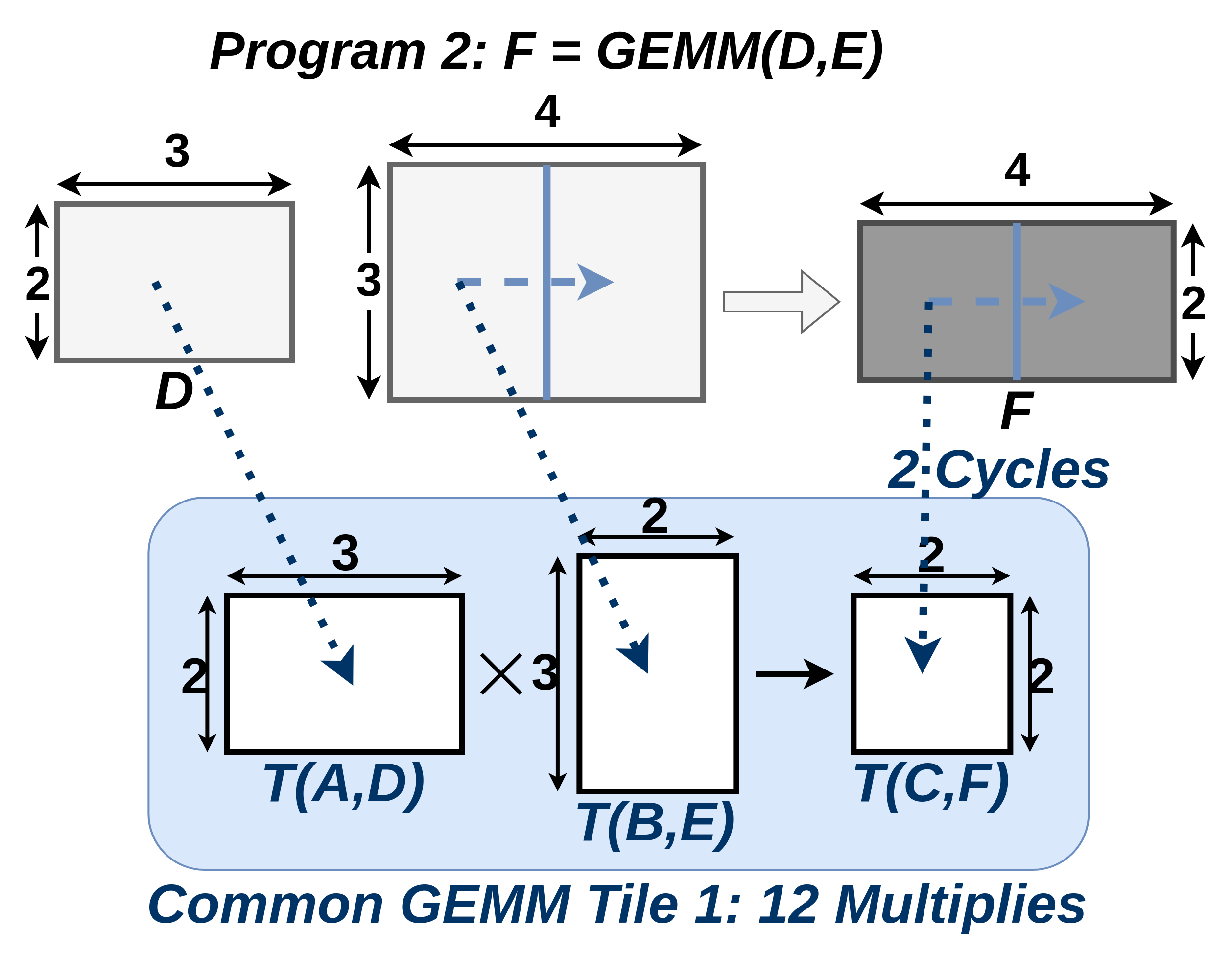}
         \caption{\centering Minimum Tiling GEMM(D,E)}
         \label{fig:tile_p2_t1}
     \end{subfigure}

     \hfill

     \begin{subfigure}[b]{0.49\linewidth}
         \centering
         \includegraphics[width=\linewidth]{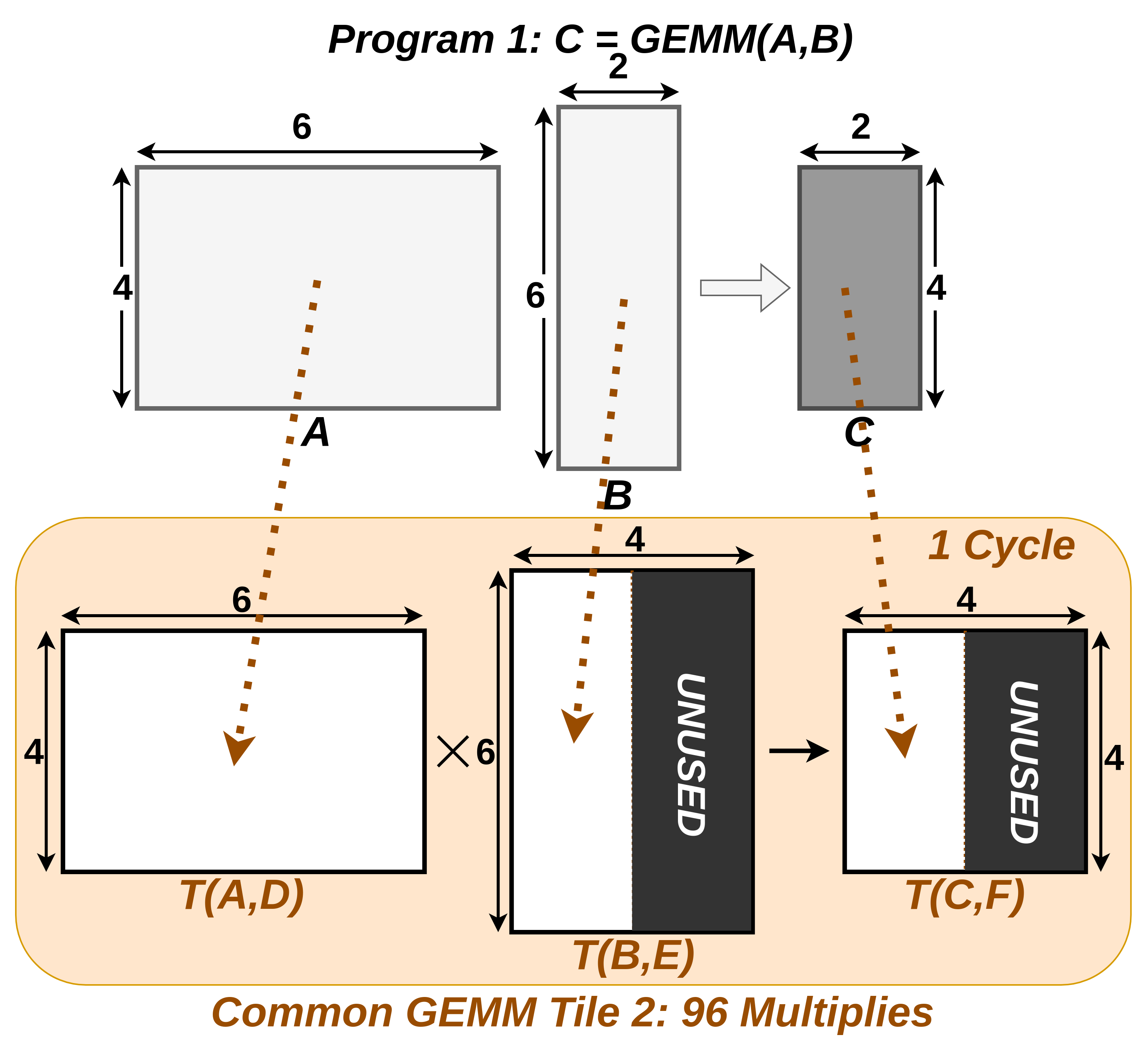}
         \caption{\centering Maximum Tiling GEMM(A,B)}
         \label{fig:tile_p1_t2}
     \end{subfigure}
     \hfill
     \begin{subfigure}[b]{0.49\linewidth}
         \centering
         \includegraphics[width=\linewidth]{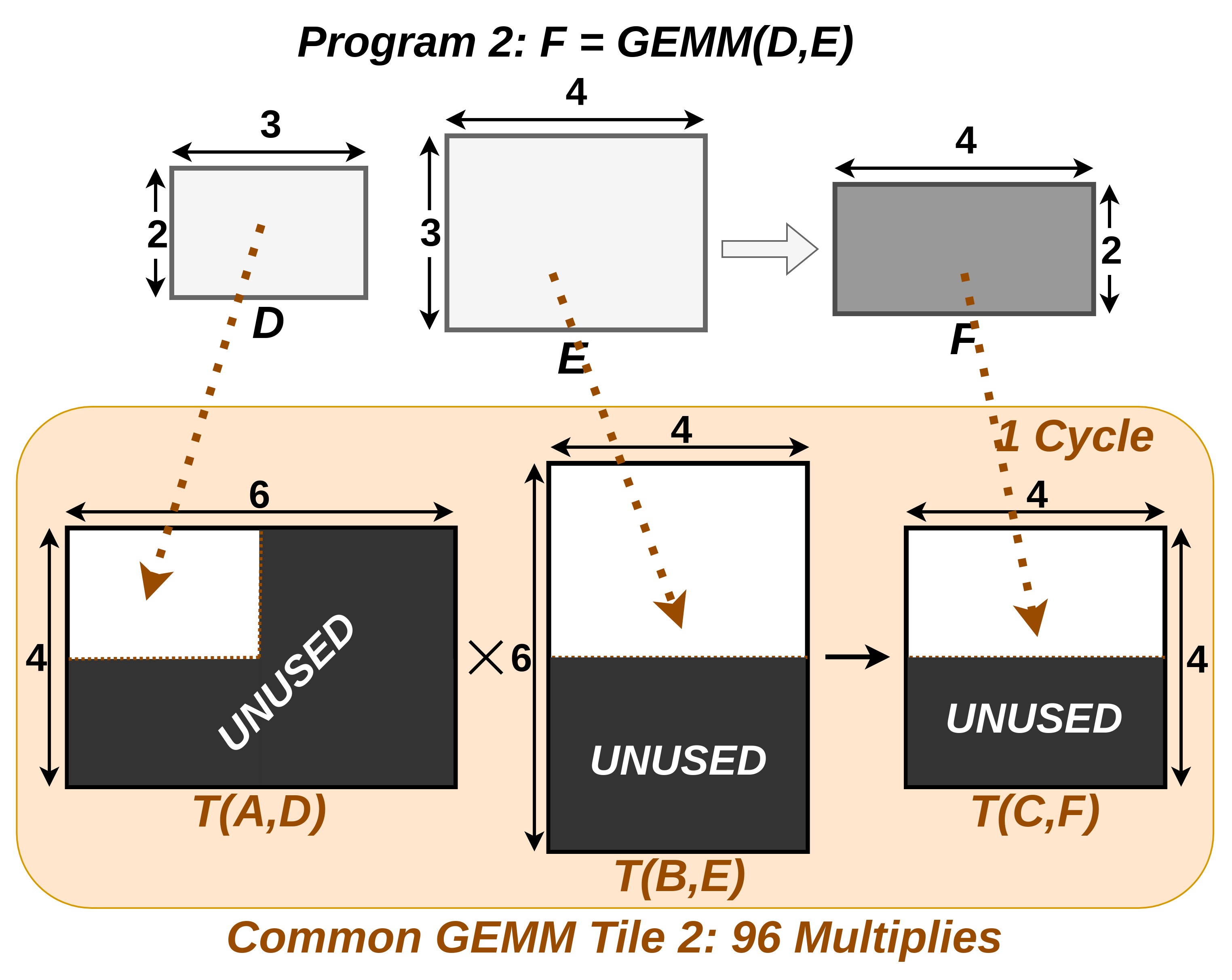}
         \caption{\centering Maximum Tiling GEMM(D,E)}
         \label{fig:tile_p2_t2}
     \end{subfigure}

     \caption{Resource/Latency trade-off with common GEMM tiles.}
    \label{fig:common_tile_tradeoff}

 \end{figure}
The second benchmark suite focuses on modularization and hardware reuse in ML HLS designs, described in detail in section \ref{sec:modular_benchmark}. We use ForgeBench first to generate 2-3 input designs per test case. The test cases are chosen with significant reuse opportunities across the input designs. We then provide the ideal modularized implementation, determined manually. These are added to the benchmark suite and serve as a reference output for future modular HLS tools. 

When deciding on the shared computational modules for an architecture, there is a tradeoff between the resource usage of the common modules and the execution time of input designs. We highlight this tradeoff with two simple matrix multiplications as an example in Fig. \ref{fig:common_tile_tradeoff}. The depicted operations $\text{GEMM}(A,B) \rightarrow C$ and $\text{GEMM}(D,E)\rightarrow F$ have dimensions (4,6,2) and (2,3,4) respectively. Both operations can be computed using a tiled GEMM unit by carefully mapping the matrices to the hardware. The first option is the \textit{minimum} tiling, highlighted in Fig. \ref{fig:tile_p1_t1} and \ref{fig:tile_p2_t1}, where the common tile is the greatest common denominator (GCD) of each dimension. This guarantees that both computations can be completed by tiling the input matrices to match the GCD dimensions and computing over multiple iterations. The GEMM operations in Fig. \ref{fig:tile_p1_t1} and \ref{fig:tile_p2_t1} take 8 and 2 iterations of the tiled computation, respectively. Alternatively, the \textit{maximum} tiling can be used, where the common tile is determined with the highest common multiple (HCM) of each dimension, pictured in Fig. \ref{fig:tile_p1_t2} and \ref{fig:tile_p2_t2}. This tile size guarantees the matrices fit into common tiles, with every program computing within 1 iteration, at the expense of under-utilized hardware. This design choice is prominent in all function/tiling reuse test cases, emphasizing the necessity of a rigorous benchmark suite to evaluate future modular HLS tools and how to handle these decisions. Within the provided test suite, test cases - Tiled GEMM-{Min/Max} - and Vec/Mtx Mult - {Dot/MMV/GEMM} explicitly represent the choices of common module extraction. 

\subsection{Extensibility}

Although ForgeBench currently spans three ML domains, it is not limited to these, as it is highly generalizable and extensible to emerging applications. First, ForgeBench incorporates numerous basic and widely used operator modules for machine learning. Many complex operators can be decomposed into these fundamental components. For example, the SwiGLU operator used in LLaMA can be decomposed into the SiLU activation and element-wise multiplication, while the SiLU itself can be decomposed into the sigmoid function and element-wise multiplication. Second, even if a new operator cannot be decomposed into existing components, users can easily incorporate them into ForgeBench with minimal effort and generate synthesizable HLS designs with a JSON configuration file. Furthermore, test cases generated by ForgeBench can be seamlessly integrated with existing HLS tools such as AutoDSE~\cite{AutoDSE} and HLSFactory~\cite{HLSFactory}. These tools can take in our auto-generated benchmarks and increase the capacity and complexity of designs by including more complicated \texttt{HLS pragmas}, paving the way for more advanced HLS tool design in the future.


\section{Results} \label{sec:results}

\subsection{Representing ML Models} \label{sec:ml_results}

\begin{table}[!t]
\centering
\resizebox{\linewidth}{!}{
\begin{tabular}{|C{1.0cm}|C{2.7cm}|C{6cm}|}
\hline
    \textbf{Class} & \textbf{Model} & \textbf{Key Operations} \\
    
    \hline
    \hline

    \multirow{8}{1.0cm}{\centering \textbf{DNNs}} & ResNet-N${\ast}$ \cite{ResNet} & Convolution, MaxPool, Matrix Add, ReLU, SoftMax, BatchNorm \\ \cline{2-3}
    & VGG${\ast}$ \cite{VGG} & Convolution, MaxPool, ReLU, SoftMax, BatchNorm \\
    \cline{2-3}
    & MobileNet-1/2/3 \cite{MobileNet, MobileNetV2, MobileNetV3} & Grouped Convolution, ReLU, ReLU6, BatchNorm, HardSigmoid, HardSwish \\
    \cline{2-3}
    & SqueezeNet \cite{SqueezeNet} & Convolution, MaxPool, Matrix Add, ReLU, SoftMax, BatchNorm \\
    \cline{2-3}
    & EfficientNet \cite{EfficientNet} & Grouped Convolution, AvgPool, Matrix Add, SiLU, Sigmoid, BatchNorm \\
    \hline
    \hline
    
    \multirow{8}{1.0cm}{\centering \textbf{LLMs}} & GPT-1/2${\ast}$ \cite{GPT} & MultiHead Attention, GEMM, LayerNorm, Dropout \\ 
    \cline{2-3}
    & LLaMA${\ast}$ \cite{LLaMA} & Grouped MultiHead Attention,\newline GEMM, RMS-Norm, SwiGLU, RoPE \\
    \cline{2-3}
    & Mistral \cite{Mistral} & Sliding Window Attention,\newline GEMM, RMS-Norm, SwiGLU, RoPE\\
    \cline{2-3}
    & Gemma-1/2 \cite{Gemma_1, Gemma_2} & Grouped MultiHead Attention,\newline GEMM, RMS-Norm, GeGLU, RoPE\\
    \cline{2-3}
    & Griffin \cite{Griffin} & GEMM, RMS-Norm, GeLU, Element-wise Matrix Multiplication, Convolution\\
    \hline
         
    \end{tabular}}
      \caption{ML models supported by our framework. We provide JSON file design examples for ML models annotated with {$\ast$}.} 
    \label{tab:supported_models}
    \vspace{5pt}
\end{table}

\begin{table}[!t]
\centering
\resizebox{\linewidth}{!}{
\begin{tabular}{|C{1cm}|C{1.4cm}|C{4.9cm}|C{1.3cm}|}
\hline
    \textbf{Class} & \textbf{Base Design} & \textbf{Configurations} & \textbf{\# Testcases} \\
    
    \hline
    \textbf{GEMM} & $x\cdot A \cdot B \cdot y$ & Vector/Matrix Dimensions, Loop Order, Unroll Factors, Computation Order & 1920 \\
    \hline
    \textbf{DNN} & Conv-BatchNorm-Activation & Feature Map \& Kernel Dimensions, Grouped, Bias, Activation Function, Unroll Factors & 2304 \\
    \hline
    \textbf{LLM} & Attention-Dropout-Norm & Attention Dimensions, Heads, Grouped, With RoPE, With Dropout, Normalization Type, Dropout Probability & 1944 \\
    \hline
         
    \end{tabular}}
    \caption{Auto-generated ML benchmark suites that are ready to use.}
    \label{tab:ml_suite}
\end{table}

\begin{table*}[!tbh]
\centering
\resizebox{\linewidth}{!}{
\begin{tabular}{|C{1.3cm}|C{3.4cm}|C{1.8cm}|C{1.8cm}|C{1.8cm}|C{1.8cm}|C{2cm}|C{2cm}|C{2.2cm}|}
\hline

    \multirow{2}{1.3cm}{\centering \textbf{Test Suite}} & 
    \multirow{2}{3.4cm}{\centering \textbf{Test Case}} & 
    \multicolumn{4}{C{8.3cm}|}{\centering \textbf{Resource Utilization (LUT\%, DSP\%) Before Modularization}} & 
    \multicolumn{2}{C{4.6cm}|}{\textbf{Resource Utilization (LUT\%, DSP\%) After Modularization}} &
    \multirow{2}{2.2cm}{\centering \textbf{Change\,(\%, \%)}} \\ 
    
    \cline{3-6} \cline{7-8}

    & & \textbf{P1} & \textbf{P2} & \textbf{P3} & \textbf{Total} & \textbf{Shared} & \textbf{Total} & \\ 
    
    \hline
    \hline
    
    \multirow{7}{1.3cm}{\centering \textbf{GEMM}} & 
    Tiled GEMM - Min $\dagger$ &
    \multirow{2}{1.8cm}{\centering (12.07, 15.24) } & 
    \multirow{2}{1.8cm}{\centering (6.35, 5.08) } & 
    \multirow{2}{1.8cm}{\centering (18.18, 20.32) } & 
    \multirow{2}{1.8cm}{\centering (36.60, 40.64) } & 
    (4.03, 0.63) &
    (8.57, 0.63) &
    (-76.6, -98.45) \\
    \cline{2-2} \cline{7-9}
    
    & Tiled GEMM - Max $\dagger$ & & & & &  
    (42.95, 81.27) & 
    (79.98, 81.27) & 
    (118.5, 99.98) \\
    \cline{2-9}
    
    & Vec/Mtx Mult - Dot $\dagger$ $\ast$ & 
    \multirow{3}{1.8cm}{\centering (45.34, 81.27) } & 
    \multirow{3}{1.8cm}{\centering (6.75, 10.16) } & 
    \multirow{3}{1.8cm}{\centering (1.39, 0.63) } & 
    \multirow{3}{1.8cm}{\centering (53.48, 92.06) } & 
     (0.06, 0.63) & 
     (4.39, 1.27) & 
     (-91.79, -98.62)\\
   \cline{2-2} \cline{7-9}
    
    & Vec/Mtx Mult - MMV $\dagger$ $\ast$& & & & & 
    (1.58, 5.08) & 
    (11.9, 10.16) & 
    (-77.75, -88.96) \\
    \cline{2-2} \cline{7-9}

    & Vec/Mtx Mult - GEMM $\dagger$ $\ast$ & & & & & 
    (36.45, 81.27) & 
    (58.03, 81.27) & 
    (8.51, -11.72) \\
   \cline{2-9}
    
    & i-j-k Orders $\ast$ & 
    (42.65, 81.27)& 
    (44.35, 81.27)& 
    (43.63, 81.27)& 
    (130.6, 243.8)&  
    (32.71, 81.27)&  
    (61.42, 81.27)&  
    (-52.98, -66.67)\\

    \cline{2-9}
    
    & Vector Transpose $\ast$ & 
    (6.66, 10.16)& 
    (4.92, 10.16)& 
    -- & 
    (24.28, 20.32)&  
    (1.99, 10.16)&  
    (17.62, 10.16)&  
    (-27.42, -50)\\

    \hline
    \hline
    
    \multirow{3}{1.3cm}{\centering \textbf{DNN}} & Activation Functions $\ddagger$ $\ast$ & 
    (2.36, 0.12) & 
    (2.51, 0.12) & 
    (2.07, 0.16) & 
    (6.94, 0.4) &  
    (0.24, 0.12) &  
    (3.01, 0.24) &  
    (-56.63, -40) \\
    \cline{2-9}
    
     & Tiled Convolution $\dagger$ $\ast$ & 
    (10.97, 20.32) & 
    (18.40, 20.32) & 
    (18.44, 20.32) & 
    (47.81, 60.96) & 
    (21.95, 20.44) & 
    (26.93, 20.63) &
    (-43.67, -66.15) \\
    
    \cline{2-9}
    

    & DNN Blocks $\ddagger$ & 
    (24.62, 20.36) & 
    (23.23, 20.44) & 
    (67.26, 61.07) & 
    (115.1, 101.9) &  
    (40.50, 41.32) & 
    (91.06, 81.98) &  
    (-20.89, -19.52) \\

    \hline
    \hline

    \multirow{3}{1.3cm}{\centering \textbf{LLM}} & Tiled Attention $\dagger$ $\ast$ & 
    (7.16, 1.31) & 
    (7.14, 1.31) & 
    -- & 
    (14.3, 1.31) &  
    (5.88, 1.31) &  
    (7.99, 1.31) &
    (-44.13, -50) \\
    
    \cline{2-9}
    
    & Functional Attention $\ddagger$ & 
    (21.87, 3.25) & 
    (23.49, 2.62) & 
    -- & 
    (45.36, 5.87) &  
    (17.77, 1.99) &  
    (25.69, 3.89) &  
    (-43.36, -33.73) \\

    \cline{2-9}
    
    & LLaMA/GPT Transformers $\ddagger$ & 
    (33.63, 9.44) & 
    (15.30, 7.94) & 
    -- & 
    (48.93, 17.38) &
    (8.77, 6.23) &  
    (36.72, 9.6) &  
    (-24.95, -44.76) \\

    \hline
         
    \end{tabular}}
    \caption{ Resource usage of hardware test cases, annotated according to the available reuse $\dagger$: tiling reuse, $\ddagger$: functional reuse, $\ast$: arithmetic reuse. }
    \vspace{-15pt}
    \label{tab:modularization_results}
\end{table*}

Although different machine learning models can vary significantly in structure and data flows, most share common functional components. By defining the key operator modules in ForgeBench, described in section \ref{sec:supported_ops}, and specifying different sequences of module calls in the JSON file, we can generate hardware designs for various DNNs and LLMs. Table~\ref{tab:supported_models} summarizes the supported models and their key operators.

Furthermore, with ForgeBench, we generate an ML-focused test suite with 2304 DNN test cases, 1920 GEMM test cases, and 1944 LLM test cases. Each test case takes a baseline design and derives similar implementations with varying parameters, summarized in table \ref{tab:ml_suite}. The base DNN design is a convolution kernel, BatchNorm, and arbitrary activation function. The derivative test cases modify the convolution dimensions, parameters such as grouping or bias, unrolling the in/out channels, and the choice of activation function. The base GEMM design is a typical linear algebra operation, $x\cdot A \cdot B \cdot y$, where $x,y$ are vectors and $A,B$ are matrices. The test cases modify the dimensions, the unroll factor and order of the nested loops, and the order of operations. The LLM-focused testuite is the fundamental attention layer followed by an optional dropout layer and a normalization layer. We tune the input matrix and attention dimensions, the number of heads, the number of groups, the dropout probability, and the use of either LayerNorm or RMSnorm. While this test suite can directly be used to benchmark tools, it further highlights the ease of generating ML-oriented HLS test cases using ForgeBench, enabling user to create benchmarks focused on their unique use cases.

\subsection{Benchmark Suite for Modularized HLS} \label{sec:modular_benchmark}

In this section, we leverage ForgeBench to produce a set of benchmarks specific to module reuse. The benchmarks contain GEMM, DNN, and LLM-focused test suites. Each suite contains test cases with up to 3 input programs with varying types of hardware reuse. We explore three reuse scopes across the ML domains:

\begin{itemize}
    \item \textbf{Tiling}: Computationally intensive operations are computed using many iterations of smaller similar operations.
    \item \textbf{Functional}: Operations call the same functions in different order/amount of times.
    \item \textbf{Arithmetic}: Operations have different hardware implementations but are mathematically equivalent.
\end{itemize}

Table \ref{tab:modularization_results} presents each test case, the targeted reuse, and the computational (DSP, LUT) resource utilization targeting at ZCU102 FPGA using Vitis HLS 2024.1. We report the resource utilization of the provided modularized implementations to highlight the necessity of a set of modular HLS tools. We do not present the delay or power of each design, as without a sufficient hardware architecture and control logic to schedule and execute the input designs, they have little meaning. 

Details of each test case across the suites are described:
\paragraph{GEMM Testsuite}
The first test case in this suite involves learning a standard GEMM tile across three programs each, computing a GEMM with bias of different dimensions, which can be identified. The programs P1--3 have dimensions (96,512,128), (128,256,64), (256,128,192) respectively. Correspondingly, the minimum and maximum tiles are (32,128,64) and (256,512,192), respectively. Each of these tilings is considered a separate test case. The second test case includes programs computing a GEMM, a vector-matrix product, and a dot product. Either of the three products can be used to compute the others, leading to a range of resource utilization, with a common dot product using the least, summarized in the table in the three Vec/Mtx Mult Testcases. Identifying this sharing requires both tiling the loops and some arithmetic equivalences. The `i-j-k' Orders test case requires recognizing the GEMM equivalence over three permutations of `i-j-k'. The common GEMM block may implement any arbitrary ordering of the three loops; the provided modularized code uses `i-k-j'. The final case tests for reuse across a column/row vector-matrix product, which is equivalent under a transpose of the matrix. 

\paragraph{DNN Testsuite} 
The first test case in the DNN testuite involves learning the shared arithmetic operators from different activation functions. The programs P1--3 implement activation functions: Sigmoid, Tanh, and Exponential Linear Unit (ELU), all sharing the exponent operator. In the second test case, P1--P3 correspond to three different $3 \times 3$ convolutions, each characterized by the configuration [input channels, output channels, height, width]. Specifically, P1 = [64, 64, 14, 14], P2 = [128, 128, 7, 7], and P3 = [128, 128, 14, 14]. The minimum tiling within these configurations is a [64, 64, 7, 7] convolution kernel. As a result, P1 and P2 can be computed in four iterations each, and P3 in sixteen iterations. In the third test case, we represent convolution blocks in different DNNs using [kernel size, channels]. P1 is a convolution block in VGG-19~\cite{VGG}, comprising four repeated [$3 \times 3$, 256] convolutions with ReLU activation, followed by max pooling. P2 is from ResNet-18~\cite{ResNet} and includes two repeated [$3 \times 3$, 256] convolutions with BatchNorm and ReLU. P3 is from ResNet-50~\cite{ResNet}  and consists of [$1 \times 1$, 64], [$3 \times 3$, 64], and [$1 \times 1$, 256] convolutions, along with BatchNorm and ReLU. Both P2 and P3 also include residual connections. The modular implementation shares the [$3 \times 3$, 64] convolution, BatchNorm and ReLU operations across these three programs with correct tiling.

\paragraph{LLM Testsuite}
In this test suite, the first case tests tiling two multi-head attention configurations: P1 with 16 attention heads and P2 with 4 attention heads. The modularized implementation uses the 4-head configuration as a shared computational module. Consequently, P1 requires four iterations to complete its computation. The second test case focuses on the functional decomposition of attention. P1 is the multi-head attention module from GPT, whereas P2 is the grouped multi-head attention module from LLaMA with RoPE encoding. These attentions can be decomposed into smaller computational components, revealing that both incorporate GEMM and SoftMax submodules, which can be shared and reused. In the third test case, P1 and P2 denote the GPT and LLaMA transformer blocks, respectively. Despite GPT using LayerNorm and GeLU as its activation function and LLaMA employing RMSNorm and SwiGLU, the multi-head attention and feed-forward network components can be shared across both architectures. 


\section{Conclusion \& Future Work}


In this work, we introduce \textit{ForgeBench}, a benchmark suite tailored explicitly for next-generation HLS tools focused on machine learning architecture design. ForgeBench is accompanied by a user-friendly and highly extensible framework for automated HLS design generation and evaluation. We also provide two ready-to-use benchmarks; one includes over 6,000 representative ML test cases, and the other includes modular HLS test cases with possible resource sharing manually implemented to highlight the necessity of designing architecture-oriented HLS tools.

Existing research and recent advancements strongly support our vision for future architecture-oriented HLS tool design. Notably, equivalence-graphs (e-graphs) have shown promise for extracting shared structures across programs~\cite{BABBLE}, and have already proved effective at representing and optimizing hardware designs ~\cite{ROVER, SEER}. Consequently, we envision future HLS tools leveraging e-graphs to automatically identify common functional modules across HLS designs. After extracting common modules, the tools will generate a flexible finite-state machine and embed scheduling information into instructions that govern state transitions, enabling flexible dataflows to reuse functional modules. Furthermore, we foresee that HLS tools will perform design space exploration during architecture generation to determine the optimal module reuse choice, balancing area, and latency in a holistic and automated manner. Given these developments, and considering that ForgeBench is fully open-sourced and easily extensible, we believe it will substantially benefit ongoing and future HLS tools design and implementation.

\bibliographystyle{IEEEtran}
\bibliography{IEEEabrv,references}

\end{document}